# Deep-learning-based recognition of multi-singularity structured light

Hao Wang,[1,2,†] Xilin Yang,[3,†] Zeqi Liu,[1,2] Yijie Shen,[4] Jing Pan,[1,2] Yuan Meng,[1,2] Zijian Shi,[1,2] Zhensong Wan,[1,2] Hengkang Zhang,[1,2] Xing Fu,[1,2,*] Qiang Liu[1,2,*]

[1]Key Laboratory of Photonic Control Technology (Tsinghua University), Ministry of Education, Beijing 100084, China

[2]State Key Laboratory of Precision Measurement of Technology and Instruments, Department of Precision Instrument, Tsinghua University, Beijing 100084, China

[3]Electrical and Computer Engineering Department, University of California, Los Angeles, California 90095, United States

[4]Optoelectronics Research Centre, University of Southampton, Southampton SO17 1BJ, UK

**Abstract:** Structured light with customized complex topological pattern inspires diverse classical and quantum investigations underpinned by accurate detection techniques. However, the current detection schemes are limited to vortex beam with simple phase singularity. The precise recognition of general structured light with multiple singularities remains elusive. Here, we report a deep learning (DL) framework that can unveil multi-singularity phase structures in an end-to-end manner after feeding only two intensity patterns upon beam propagation captured via a camera, thus unleashing intuitive information of twisted photons. The DL toolbox can also acquire phases of Laguerre-Gaussian (LG) modes with single singularity and other general phase objects likewise. Leveraging this DL platform, a phase-based optical secret sharing (OSS) protocol is proposed, which is based on a more general class of multi-singularity modes than conventional LG beams. The OSS protocol features strong security, wealthy state space and convenient intensity-based measurements. This study opens new avenues for vortex beam communications, laser mode analysis, microscopy, Bose-Einstein condensates characterization, etc.

**Keywords:** structured light; vortex beams; orbital angular momentum; deep learning; optical secret key sharing

## 1 Introduction

In addition to linear momentum, photons can carry angular momentum divided into two forms, spin and orbital [1,2]. The spin angular momentum usually refers to light's polarization [3], and the orbital angular momenta (OAM) arises if the wavefront is helically shaped with a phase singularity on axis [4]. The typical unit of the helical phase is the azimuthal dependence $e^{i\ell\varphi}$, where $\ell$ reveals topological charge (TC) and $\varphi$ is the azimuthal angle [5]. In contrast to conventional vortex beams with single singularity, e.g. Laguerre-Gaussian (LG) and Bessel modes, recent advances highlight the general structured light with distributions of multiple singularities, such as singularity array [6,7], vortex lattices [8], and SU(2) vortex geometric modes [9-11]. Vortex beams have triggered intense research over the past three decades [12] for their potentials in both classical (optical tweezers [13,14], holographic encryption [15], communications [16], measurements [17,18], nonlinear effects [19,20]) and quantum applications (quantum secret sharing [21], quantum switches [22], quantum information processing [23], quantum states tomography [24]). For most of these pertinent application scenarios using optical vortices, accurate detection of OAM is a basic premise, which may resort to interference [25], diffraction [26], log-polar transformation [27], spiral transformation [28], multiplane light conversion [29], and deep learning algorithms [30], to name a few.

However, the current detection methods mainly work for single-singularity vortices. The precise measurement of more complex multi-singularity counterparts still remains a demanding task. This task requires detecting all vortices first and quantifying each vortex then. On the other hand, all the topological information of optical vortices is stored in phase whereupon a relevant issue arises: can we decode the phase directly? Indeed, phase recovery in the presence of singularities is a long-standing bugbear [31,32]. Researchers have invoked transport of intensity equation (TIE) [33], Gerchberg-Saxton based algorithm [34], wavefront modulation [35,36], diffractive imaging [37-39], or interferometer [40,41], to retrieve the phases of vortex beams. However, these profound endeavors still suffer from requiring multiple high-precision (~µm) axial intensity measurements and rigorous boundary conditions, or non-intuitive iterative algorithms, or much *a priori* knowledge, or being vulnerable to experimental noises. Most importantly, many methods are feasible for simple vortices only and the recovered performances were unfulfilling.

In this work, we propose a deep learning framework, which is dubbed "VortexNet", to overcome these hurdles. It's mention-worthy that recent years have witnessed transformative advances of deep neural networks in analyzing vortex light, spanning from scalar [30,42-48] to vectorial ones [49]. Most previous achievements regard single-singularity OAM recognition as a classification or a regression task. The outputs of their networks are only simple numbers, which relate to the exact TCs. However, we here handle it as a generative problem whereby VortexNet yields complex phases containing dozens of singularities in an end-to-end manner, thus unleashing rich and intuitive physics information to comprehend twisted photons. Here we demonstrate the phase reconstruction approach onto a set of multi-singularity SU(2) vortex modes, which accommodate conventional LG vortex modes as simple members and characterize high-dimensional topological properties [50]. Benefitting from the satisfactory performance of VortexNet, a phase-based optical secret sharing (OSS) protocol is demonstrated as an exemplary application. VortexNet can acquire phases of SU(2) beams with mode accuracy up to 93.6% from only two measurements of intensity. It also works for LG modes even with large TCs and degenerate intensities as well as other general phase objects. Therefore, it is a universal and practical tool for many classical and quantum information processing systems.

## 2 Methods

As a typical example of complex vortex beams with multiple singularities, SU(2) mode can be excited in a special resonator [51] and it can be launched by coherently superposing a set of LG modes,

$$\psi_{Q,n_0,M}(x,y,z;\phi) = S(\psi_{p,\ell}^{LG}), \quad (1)$$

where $\psi_{p,\ell}^{LG}$ represents LG mode with radial index $p$ and angular index $\ell$. The function $S(*)$ is defined to describe SU(2) superposition rule briefly, integer $Q$ refers to how many folds in the rotational symmetry, $n_0$ decides the TC of axial vortex, $M$ reveals its localized OAM of off-axis singularities and $\phi \in (0,2\pi)$ determines initial phase. Each SU(2) vortex mode corresponds to a pair of parameters $(Q, n_0, M)$ unravelling the topological information. By utilizing the multiple parameters of SU(2) modes, researchers successfully realized the classical analogy of high-dimensional quantum Greenberger-Horne-Zeilinger states recently [50]. Figure 1 illustrates the difference between circular shaped LG mode $(p, \ell) = (0,8)$ whose intensity always remains in circular symmetry, and SU(2) mode $(Q, n_0, M) = (4,8,9)$ which is favored with large OAM, multiple singularities, helical star-shaped pattern and spiral propagation trajectory (See Supplemental Material I for more details of SU(2) modes).

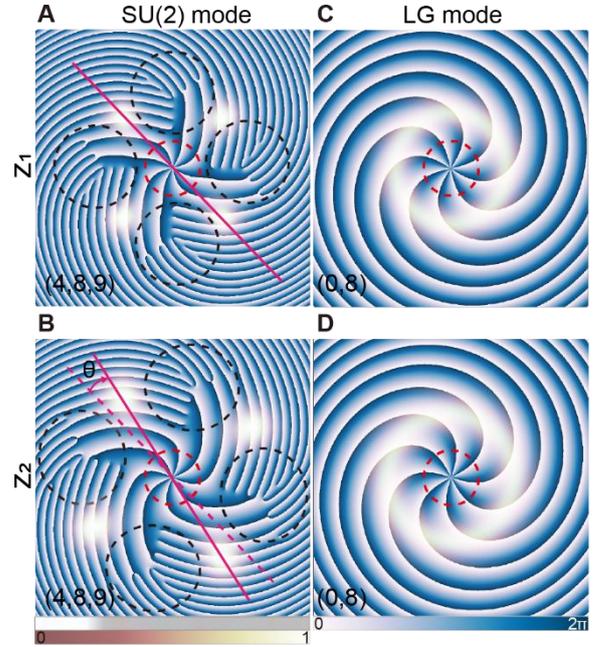

**Figure 1:** Comparison of SU(2) vortex mode (A)-(B) and LG mode (C)-(D) at two axial positions. Other than the single singularity of LG mode induced by central OAM as indicated by red circles, SU(2) mode characterizes more singularities shown by black circles. Besides, as SU(2) mode propagates from $z_1$ (a) to $z_2$ (b), its intensity undergoes an azimuthal shift $\theta$ while that of LG mode remains circularly symmetric. Note the intensity layers are set with opacity.

Inspired by TIE [32,52], a prevailing benchmark model in phase retrieval assignments, wherein the phase at the target plane can be unwrapped given the target-plane intensity and intensity differential, we construct VortexNet to calculate phases based on *two* intensity patterns $I_1$ and $I_2$ as elucidated in Figure 1: one is that of target plane,

$$I_1 = \left|\psi_{Q,n_0,M}(x,y,z_r;\phi)\right|^2, \quad (2)$$

where $z_r$ is the Rayleigh distance. The other is the intensity pattern of the defocused plane,

$$I_2 = \left|\psi_{Q,n_0,M}(x,y,z_r+\Delta z;\phi)\right|^2, \quad (3)$$

where $\Delta z$ is the diffraction distance after the target plane. In the experiment, the defocused light can be characterized by

$$\psi_{Q,n_0,M}(x,y,z_r+\Delta z;\phi) = \frac{\exp(ik\Delta z)}{i\lambda\Delta z}\exp\left[i\frac{k(x^2+y^2)}{2\Delta z}\right]$$

$$F\left\{\psi_{Q,n_0,M}(x_0,y_0,z_r;\phi)\exp\left[i\frac{k(x_0^2+y_0^2)}{2\Delta z}\right]\right\}_{f_x=\frac{x}{\lambda\Delta z},f_y=\frac{y}{\lambda\Delta z}}, \quad (4)$$

where $k = 2\pi/\lambda$ is the wavenumber, $F(*)$ denotes the spatial Fourier transformation. As a notable example, we plot these two experimental patterns of mode $(Q,n_0,M) = (5,6,7)$ in the inset of Figure 2. With visible $I_1$ and $I_2$ as inputs at hand, VortexNet is expected to reconstruct the invisible phase,
$$P_{Q,n_0,M}(x,y,z_r;\phi) = \arg[\psi_{Q,n_0,M}(x,y,z_r;\phi)], \quad (5)$$
where $\arg(*)$ returns the argument of a complex variable.

To train the VortexNet, massive data samples are necessary. Rather than utilizing a laser cavity, which is challenging for efficient on-demand modes generation, we collect datasets via spatial light modulator (SLM) as illustrated in Figure 2. More specifically, a 532 nm laser (CNI Laser, MGL-III-532) is employed as the source of Gaussian beam. After being expanded eightfold, it undergoes modulation of the SLM (Hamamatsu, X13138-04, resolution of 1280×1024, pixel size of 12.5 μm). The SU(2) state $\psi_{Q,n_0,M}(x,y,z_r;\phi)$ is encoded into the computer-generated-holographic mask through complex amplitude modulation technique [53,54]. Two lenses L1 (150 mm) and L2 (75 mm) constitute a 4-f system, through which the conjugate image of target vortex mode is recorded by the CMOS camera (AVT Mako G-131B, resolution of 1280×1024, pixel size of 5.3 μm), which completes our first measurement. Then the CMOS is moved backward for 10.00 mm through a one-dimensional translation stage (GCM-830303M, resolution of 0.01 mm), leading to a weak diffraction of the SU(2) mode, the intensity of which is recorded as the second measurement.

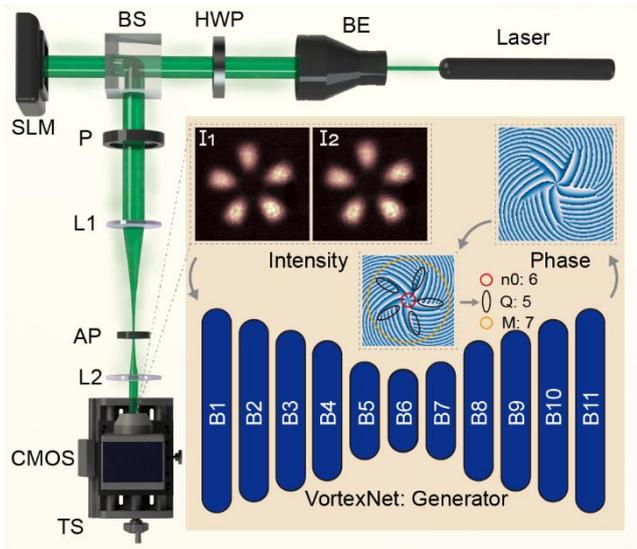

**Figure 2:** Schematic of experimental layout for the SU(2) mode creation and vortex phase acquisition. BE: beam expander; HWP: half-wave plate, which aligns the source to be p-polarized; BS: beam splitter, which assures the optical structure is rectangular; SLM: spatial light modulator; P: polarizer, which adjusts the light intensity; L1, L2: lens; AP: aperture, which filters out the first diffraction order; CMOS: camera; TS: translation stage, which moves the CMOS in $z$ direction. After recording two intensity patterns with a longitudinal interval of 10.00 mm, the generator of VortexNet reconstructs the phase and then the value of the parameters $(Q, n_0, M)$ can be decoded accordingly. B: block, configurable unit of the generator part of VortexNet.

VortexNet is inherently an adapted conditional generative adversarial network (cGAN) (See Supplemental Material III for network details), containing a generator and a discriminator [55,56]. GANs are an instrumental category of deep neural networks with ever-improving results that can extract the hierarchical features of input data and enhance the prediction accuracy [57]. In this case, the generator outputs a "fake phase" and this output along with labelled phase (true phase) are both fed into the discriminator. The discriminator strives to evaluate whether the input is real or fake. The two networks are trained jointly and they reach the "Nash equilibrium" when the training ends [58]. At this moment, the discriminator struggles to distinguish true or false and the generator returns a convincing phase.

## 3 Results

The network is first trained using an experimental dataset with mode parameters $Q \in \{3,4,5,6\}$, $n_0 \in \{1, 2, \cdots, 10\}$ and $M \in \{1, 2, \cdots, 10\}$ (See Supplemental Material II.A for dataset acquisition details). For each set of parameters $(Q, n_0, M)$, the initial phase ranges from 0 to $1.96\pi$ with an incremental value of $0.04\pi$. There are overall 400 different states and 20000 pairs of intensity images within the experimental dataset. Among them, we select 86% for training, 10% for validating and 4% for testing the VortexNet. Some blind test results are shown in Figure 3(A)-(D), showing excellent agreement between the ground truth and network output (See more in Supplemental Material II.B). Once the generator performs inference of the phase, one can decode the carried information of SU(2) mode. Based on three phase read-out rules defined in Supplemental Material I, the recognition accuracies of parameters $Q$, $n_0$ and $M$ are summarized in the confusion matrixes of Figure 4. The *accuracy* refers to the proportion of correctly classified modes and is calculated by comparing the reconstructed phase with respect to its corresponding ground truth. Accuracies of both $Q$ and $M$ reach 100% while $n_0$ achieves 93.6%. Hence the overall mode accuracy of 93.6% is already obtained. To further evaluate the quality of generated phase structures quantitatively, we calculate the peak-signal-to-noise (PSNR), structural similarity index (SSIM) and image correlation coefficient (CC) metrics. The average values of them are 14.89 dB, 0.75 and 0.76 respectively (See Supplemental Material II.B for details). These values are not that prominent because of nonideal experimental conditions like power fluctuation of laser source, lens glare, and CMOS noise. These imperfections are incorporated into the training process to beset the metrics of reconstructed phase. However, thanks to the read-out rule, some local flaws may degrade three metrics, they don't mislead us to identify the mode parameters and the mode accuracy of 93.6% is still acceptable. We note that with an improved system, the results can be enhanced. To validate this claim, we retrained a VortexNet based on exclusively simulated intensities and the mean PSNR, SSIM and CC are improved to be 16.46dB, 0.84 and 0.85. Remarkably, the recognition accuracy of $(Q, n_0, M)$ reaches 100%.

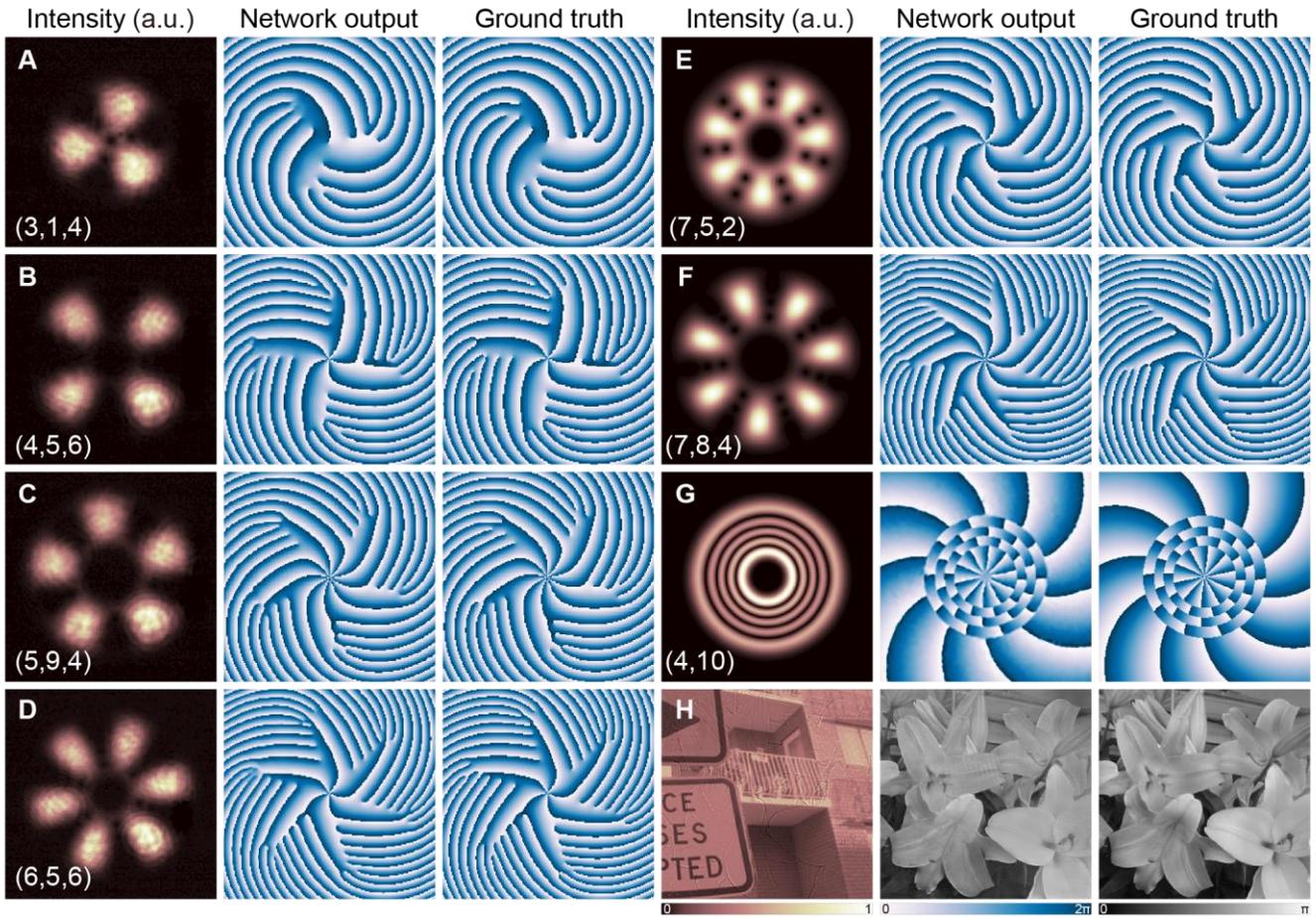

**Figure 3:** Inference results of VortexNet on experimental test set (A)-(D), modes out of training, validation and test set (E)-(F), LG modes (G) and general phase objects (H). First and fourth column: normalized intensity; Second and fifth column: VortexNet output; Third and sixth column: phase ground truth. More results are available in Supplementary Materials B.

Later, to confirm VortexNet is learning to transform extracted features into phase structures rather than memorizing all the modes and overfitting the training set, we establish the second test set containing 64 states where $Q = 7, n_0 \in \{1,2,...,8\}, M \in \{1,2,...,8\}$, which separate from training, validation and the first test set. We let the trained model based on digitally generated dataset to infer. Two results are depicted in Figure 3(E)-(F) with the overall accuracy reaching 93.75% and the average PSNR, SSIM, CC being 12.00 dB, 0.57, and 0.58.

In order to analyze the physics-informed training phase of VortexNet, i.e. why two intensity measurements are inevitable, we naturally train a model with only $I_1$ as input. It's a strongly ill-posed inverse problem to retrieve $P_{Q,n_0,M}(x, y, z_r; \phi)$ and indeed the poor blind testing results in Supplemental Material II.C bear this out. Interestingly, we train another VortexNet akin to TIE, i.e. we now input $I_1$ and a differential approximated as $I_{diff} = (I_1 - I_2)/\Delta z$. The convergence time and mode accuracy result are similar to that of $I_1$ and $I_2$ as input. It's quite reasonable because a linear operation does not minish the amount of information and is easily learnable for deep neural networks.

In addition, we realize that this computational network not only can be applied for multi-singularity vortex modes, but also more universal phase reconstruction missions. Part of the LG modes results as well as other general phase images based on simulated datasets are illustrated in Figure 3(G)-(H) (See more in Supplemental Material II.D). It's notable that VortexNet trained with proper dataset can effectively solve the intensity degenerate problem that previous endeavors encountered [30,43], where the intensities of $\psi_{p,\ell}^{LG}$ and $\psi_{p,-\ell}^{LG}$ are identical to make previous network unable to distinguish.

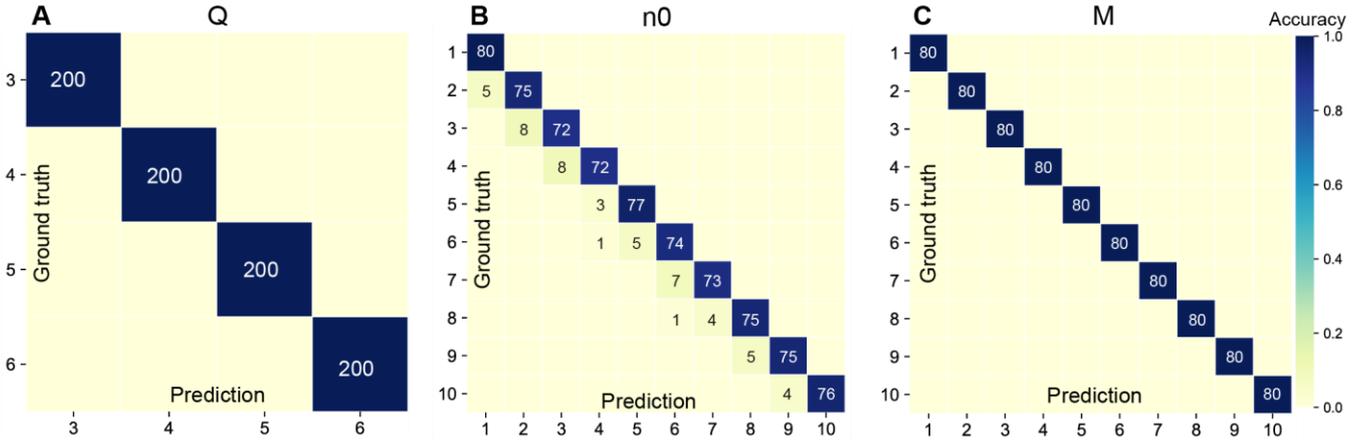

**Figure 4:** The confusion matrixes of experimental testing results, indicating the number of correctly (diagonal) and incorrectly (off-diagonal) predicted mode parameters (A) $Q$, (B) $n_0$ and (C) $M$. Note the color of every box reflects the normalized accuracy.

As we discussed before, VortexNet can directly deliver phases of SU(2) modes through indiscernible intensities hence it opens new pathways for utilizing high-dimensional topological properties in information dissemination. As data security becomes increasingly crucial, optical-based secret sharing thrives due to its abundant degrees of freedom, broad bandwidth and susceptibility to eavesdropping [59,60]. We, therefore, devise a phase-based OSS protocol enabled by VortexNet as depicted in Figure 5.

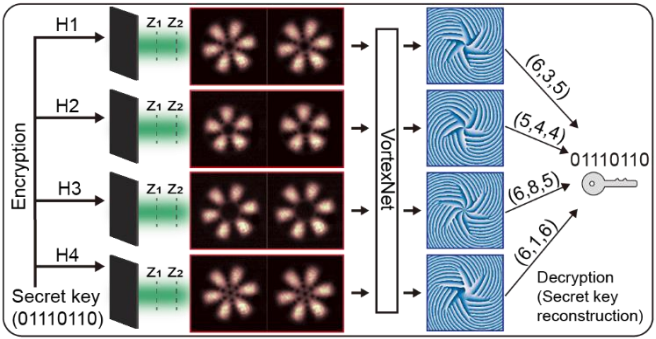

**Figure 5:** The phase-based OSS workflow. The secret key is encrypted into holograms (H1, H2, H3 and H4) that modulate the incoming light into distinct SU(2) modes. Each shareholder catches two intensities of SU(2) mode at position $z_1$ and $z_2$. The decryption must involve honest collaboration and VortexNet to calculate the phases.

First we allocate every mode $(Q, n_0, M)$ to an 8-bit binary number (See Supplemental Material IV for detailed rules). Suppose the secret $S$ is encoded as an 8-bit binary number (01110110 in this case) and in a traditional secret sharing scheme, the distributor will give three of four shareholder each a random 8-bit number ($N_1, N_2$, and $N_3$). The last player has to get the result of $S \oplus N_1 \oplus N_2 \oplus N_3$, where $\oplus$ denotes bitwise exclusive or (XOR). Instead, to complicate the message, we allot a bundle of SU(2) light to every player. Based on the experimental scheme in Figure 2, we implement the allocation operation through four holograms and in fact this can also be implemented technically via one hologram that customizes four diffraction orders of further [61]. Every player takes two shots of in incoming light coming SU(2) beam as his/her share of the secret. To decrypt the key, firstly all shareholders need to interpret two pieces of intensities through VortexNet, which is held in the hands of "the president", who will then honestly take out their shares based on retrieved phases and adopt bitwise XOR operation. The high security stems from exotic and noninterpretable SU(2) intensity patterns, wealthy state space and additional layer endowed by VortexNet.

## 4 Discussions and conclusions

The general significance of VortexNet covers several aspects: it is envisaged to recover phases of other multi-singularity beams such as vortex lattice [62] as well as diverse categories of structured light——LG beam, Airy beam and Bessel beam, to name a few——it liberates us by encoding/decoding *phase* information from only caring about intensity utilizations of shaped beams. The network structure can also be cascaded with another GAN which accommodates VortexNet's phase as input and gives mode parameters directly [63]. This way the detection is reduced to be like Refs. [30,49]. In experiments, the intensity interval is fixed to be 10.00 mm, which doesn't pose rigorous conditions for implementation. And the network is robust against small misalignment. The adverse atmosphere turbulence on structured light may be mitigated by an improved VortexNet without recourse to a probe beam [64] in future research. In addition, VortexNet can also be enforced on complex amplitude reconstruction assignments, which can aid in digital holographic imaging [65]. Besides, the OSS is scalable both in terms of SU(2) modes number and secret-splitting rule. The first point concerns that as the neural networks nowadays progress towards bigger model architecture, bigger dataset and stronger computation power, one can scale up the SU(2) modes and construct a benchmark dataset/library for more sophisticated encryption. Secondly, the proposed OSS can be readily extended to $(t, n) - threshold$ scheme where $t$ or more of $n$ players can decipher the key, and also grafted onto classical Shamir's scheme, the Chinese remainder theorem-based scheme, etc. Furthermore, one can adopt the affluent SU(2) states to transmit information in a mode-multiplexing or shift-keying manner [66,67]. Notably, when a frequency-degenerate resonator emits SU(2) modes in the lab, one has to blindly select $(Q, n_0, M)$ in simulation to find the most similar one to characterize the emission mode due to non-interpretable intensity pattern before. Now VortexNet raises a hand to do the mode analysis job, which greatly facilitates the development of structured beam laser [68]. Last but not the least, the VortexNet-based techniques may also find further applications on electronic, X-ray as well as acoustics systems.

To conclude, we demonstrate a new approach to tackling structured beams enabled by deep learning, in particular phases with single/multiple singularities. The topological properties are revealed directly, accurately and robustly by using only two convenient intensity-based measurements as inputs, even in the presence of inherent noises and instabilities. Empowered by SU(2) vortex modes with high-dimensional quantum analogy properties and great state space, numerous schemes employing them as information carriers are feasible now and we demonstrate a novel OSS protocol as an instance. This DL-assisted platform may promise relevant implications in OAM communications [48], laser mode analysis [68], microscopy [69], Bose-Einstein condensates characterization [70], etc.

We acknowledge Jia Guo, Lin Zhang, Yan Wang and Jiading Tian. This work is supported by the National Natural Science Foundation of China (61875100) and National Natural Science Foundation of China (61975087).

*Corresponding author.
qiangliu@tsinghua.edu.cn
fuxing@tsinghua.edu.cn
†H.W. and X.Y. contributed equally to this work.